\documentstyle[11pt,aaspp]{article}
\newcommand{\postscript}[2]
 {\setlength{\epsfxsize}{#2\hsize}
  \centerline{\epsfbox{#1}}}
\def\ref#1{\par\noindent \hangindent=0.4in \hangafter=1 #1 \par}
\def\eqalign#1{\null\,\vcenter{\openup\jot \m@th
  \ialign{\strut\hfill$\displaystyle{##}$&$
     \displaystyle{{}##}$\hfill \crcr#1\crcr}}\,}
\def\tempest%
{\begin{array}{ccc}
1 & 1 & 1 \\
1 & 1 & 1 \\
4 & 3 & 8
\end{array}}
\def\gsim{{{}_>\atop{}^{{}^\sim}}}
\def\lsim{{{}_<\atop{}^{{}^\sim}}}

\begin{document}

\title{Microlensing by Multiple Planets in High Magnification Events}

\author{B. Scott Gaudi}

\affil{Ohio State University, Department of Astronomy, Columbus, OH 43210 \\
gaudi@astronomy.ohio-state.edu}

\author{Richard M. Naber and Penny D. Sackett}

\affil{Kapteyn Astronomical Institute, 9700 AV Groningen, The
  Netherlands \\
 richard, psackett@astro.rug.nl}

\begin{abstract}

Microlensing is increasingly gaining recognition as a powerful method
for the detection and characterization of extra-solar planetary
systems.  Naively, one might expect that the probability of detecting
the influence of more than one planet on any single microlensing
light curve would be small.  Recently, however, Griest \& Safizadeh
(1998) have shown that, for a subset of events, those with minimum 
impact parameter $u_{min} \lsim 0.1$ (high magnification events), 
the detection probability is 
nearly 100\% for Jovian mass planets with  projected separations in the range 
0.6--1.6 of the primary Einstein ring radius $R_E$, and remains substantial 
outside this zone.  In this Letter, we point out that this result implies
that, regardless of orientation, {\it all} Jovian mass planets with separations
near 0.6--1.6$R_E$ dramatically affect the central region of the 
magnification pattern,
and thus have a significant probability of being detected (or ruled out)
in high magnification events. 
The probability, averaged over all orbital phases and inclination angles, of 
two planets having projected separations within $0.6$--$1.6R_E$
is substantial: 1-15\% for two planets with the intrinsic orbital 
separations of Jupiter and Saturn orbiting around 0.3--1.0$M_\odot$ 
parent stars.  We illustrate by example the complicated magnification 
patterns and light curves that can result when two planets are present, 
and discuss possible implications of our result on detection efficiencies 
and the ability to discriminate between multiple and single planets in 
high magnification events.

\end{abstract}


\centerline{submitted to the {\it Astrophysical Journal Letters}: March 24, 
1998}
 
\newpage
 
\section{Introduction}

A planetary microlensing event occurs whenever the presence of a planet creates
a perturbation to the standard microlensing event light curve.
These perturbations typically have magnitudes of $\lsim 20\%$ 
and durations of a few days or less.
First suggested by Mao \& Paczy\'nski (1991) as a
method to detect extra-solar planetary systems, the possibility was
explored further by Gould \& Loeb (1992), who found that roughly 15\%
of microlensing light curves should show evidence of planetary
deviations if all primary lenses have Jupiter-mass planets with
orbital separations comparable to that of Jupiter.  
Although these probabilities are
relatively high, the use of microlensing to discover planets was
largely ignored since in order to detect the primary events the 
microlensing survey teams must monitor millions of stars in very crowded 
fields, resulting in temporal sampling
that is too low ($\sim 1\, {\rm day}$) and photometric errors that
are too high ($\gsim 5\%$) to detect most secondary planetary deviations 
(Alcock et al.\ 1997a).  

Recently, the situation has changed dramatically as the real-time 
reduction of the survey teams has enabled them to issue electronic 
``alerts,'' notification of on-going events detected before the peak
magnification (Udalski et al.\ 1994, Pratt et al.\ 1996), 
allowing other collaborations to perform 
special purpose observations of the alerted events.  These additional 
observations include denser photometric sampling by the PLANET 
and GMAN collaborations (Albrow et al.\ 1996, 1997, 1998 and Pratt 
et al.\ 1996, Alcock et al. 1997b) as well as spectroscopic follow-up 
of particular events (Lennon et al.\ 1997). 
Over 60 events are currently alerted per year towards the Galactic Bulge. 
Since only a handful of these are on-going at any given time, 
monitoring teams can sample events very densely and with high 
photometric accuracy, enabling the
detection of many second order effects, including --in principle--
planetary anomalies.  No clear planetary detections have yet been made in 
this way, but preliminary estimates 
of detection efficiencies show that PLANET, over the next two observing 
seasons, should be 
sensitive to planetary anomalies caused by Jovian planets orbiting a 
few AU from their parent star (Albrow et al.\ 1998). 
Thus, if these kinds of planets are common, they should be detected soon.  
If not, microlensing will be
able to place interesting upper limits on the frequency of such systems. 

These observational developments have been accompanied by an explosion of
theoretical work, including further studies of detection
probabilities and observing strategies incorporating a variety of
new effects (Bolatto \& Falco 1994, 
Bennett \& Rhie 1996, Peale 1997, Sackett 1997, DiStefano \& Scalzo 1998a,b), 
demonstration of planetary microlensing light curves (Wambsganss 
1997), explorations of the degeneracies in the
fits of planetary events (Gaudi \& Gould 1997, Gaudi 1998), and a study of
the relation between binary and planetary lenses (Dominik 1998).
It would thus seem that the theoretical understanding of the detection and
characterization of planetary systems using microlensing should be
well in hand.

Surprisingly, however, the field still has surprises to offer.
Recently, Griest \& Safizadeh (1998, hereafter GS) came to a rather startling
conclusion:  for microlensing events
with minimum impact parameter $u_{min} \lsim 0.1$
(maximum magnification $A \gsim $10), the detection probability is nearly 
100\% for Jovian mass planets with projected separations lying 
within $0.6$--$1.6$ of the Einstein ring of the primary, i.e., the so-called
``standard lensing zone.''  In fact, GS found 
that the detection probability for this subset of events is higher for 
{\it all} projected 
separations, and preferentially so for smaller separations.  Since the 
probability that an event will have impact parameter $u_{min} < 0.1$ is
$\sim$10\%, this means that, for $\sim$10\% of all events, the
existence of a planet in the lensing zone can be detected or ruled out.
The primary point of this {\it Letter\/} is to stress that 
the conclusions of GS necessarily imply that, 
regardless of orientation, {\it all} Jovian mass planets in the lensing
zone dramatically affect the central region of the magnification pattern,
and thus have a significant probability of being detected (or ruled out)
in small impact parameter (high magnification) events.

We present here a preliminary exploration of
microlensing by lenses orbited by multiple (two) planets.  
Because our results are
intimately tied to those of GS, we refer the reader to that paper 
for a more thorough investigation of detecting
single planets with high magnification events.  Note that we
will use high magnification here to mean events for which the minimum 
impact parameter from the primary is $u_{min} < 0.1$.
In order to assess the frequency with which multiple planets may lie at 
detectable separations, we calculate in \S 2 the
probability of two planets having projected separations 
in the standard lensing zone, and indicate why an even larger zone  
is more appropriate for high magnification events.  
In \S 3, we briefly review the formalism needed for calculating
the magnification patterns created by single, double, and
triple lenses, and in \S 4, we present sample light curves.  
In \S 5, we
discuss possible implications of our results and conclude.

\section{``Lensing Zone'' Frequencies for Multiple Planets}

The ``standard lensing zone'' is generally defined as the annular region
in the source plane with $0.6 \le r \le 1.6R_E$, where $R_E$ is the 
Einstein ring of the parent star,
$$
R_E= {\left[ 4GM D_{OL}(1-D_{OL}/D_{OS})\right]^{1/2} \over c}=
3.5\, {\rm AU} \left(M\over M_{\odot}\right)^{1/2},
\eqno(2.1)
$$
\noindent and $M$ is the mass of the primary lens.  
For the scaling relation on the far right-hand-side, we have 
assumed a source distance $D_{OS}=8\, {\rm kpc}$ and the lens 
distance $D_{OL}=6\,{\rm kpc}$.  
With these assumed distances, the lensing zone corresponds to
$2.1-5.6\, {\rm AU}$ for a 1.0$M_\odot$ primary, and $1.1-3.0\, {\rm AU}$
for a $0.3M_{\odot}$ primary.  As demonstrated by Gould \& Loeb (1992),
this zone roughly corresponds to the range of projected separations 
$b\equiv r/R_E$ for which planets will have substantial detection
probabilities, averaging over all possible events.  However, 
as we will discuss, for the subset of high magnification events, the 
relevant zone of planetary separations is somewhat more extended. 

The standard lensing zone boundaries are defined by the image 
positions for single lens microlensing for a source position equal 
to $R_E$, the largest source position for 
which current microlensing experiments will alert (magnification $A = 1.34$).  
Source positions closer to the lens will result in larger magnifications  
and image positions closer to $R_E$.  
If planetary detection is defined as the source crossing the caustic 
structure (induced by the binary lens) that lies near the planet position, 
then the planet must be close to these image positions and thus  
within the lensing zone in order to have a measurable effect.  
With such a definition 
of planetary detection, one would expect multiple planet detection 
to happen only rarely since the source trajectory 
must traverse both planetary positions, both of which 
must lie within the lensing zone.

For high magnification events ($A >10$), GS have shown 
that the planets with mass ratio 
$q\gsim 0.001$ may be detected with nearly 100\% probability even 
when they lie somewhat outside the lensing zone.  This is because the 
planetary anomaly is caused by the source approaching or crossing the 
central caustic 
(near the primary), not the planetary caustic.    
This in turn implies that the detection probabilities for 
{\it multiple planets\/} in the lensing zone will also be high, 
providing such a scenario occurs frequently enough.  

Given two planets with true
separations (in units of $R_E$) of $a_1$ and $a_2$, we thus wish to 
calculate the probability that the projected separations $b_1$ and $b_2$ 
will fall in the lensing zone.  The relation between the true and projected
separations is, 
$$
b_k=a_k[\cos^2{\phi_k} + \sin^2{\phi_k}\cos^2{i}]^{1/2},
\eqno(2.2)
$$
where $i$ is the orbital inclination, $\phi_k$ is the orbital
phase of planet $k$, and we have assumed circular, co-planar orbits.
The calculation of the probability involves a three-dimensional
integral over $\cos{i}$, $\phi_1$, and $\phi_2$, the distributions of
which are flat.  The result is shown in Fig.~1 as a
function of $a_2$, for several discrete values of $a_1$ representing 
known or plausible planetary systems with Jovian mass planets. The
separations in physical units (${\rm {AU}}$) scale according to Eq.~2.1.  

It is apparent from Fig.~1 that the probability of two Jovian 
planets falling in the lensing zone, regardless of their relative
positions, may be quite high.  Note, in particular, the long tail for
higher true separations $a_2$.  Furthermore, the {\it conditional}
probability (lower panel of Fig.~1), i.e., the probability that
both planets will be in the lensing zone given that one of the planets
already meets this criterion, is even higher.  For high magnification events,
this implies that it is highly probable that if
deviations from one planet are present, 
deviations from the second planet are present as well.
For planets with true separations equal to that of 
Jupiter and Saturn ($5.2$ and $9.5\,{\rm {AU}}$, respectively), the
probability of both planets being in the lensing zone is $14\%$ if the
planets are orbiting a $1.0M_{\odot}$ primary, and $1\%$ for a
$0.3M_{\odot}$ primary.  

Radial velocity techniques have discovered 
several Jovian-mass planets, many with orbital separations substantially 
smaller than 1~AU (Mayor \& Queloz 1995, Butler \& Marcy 1996), making them 
difficult to detect via microlensing.  Other planets detected by radial 
velocity methods, like the 3$M_J$ mass planet orbiting the G0 star 47UMa
on a circular orbit at 2.1~AU, would be detectable in high magnification 
events, especially if in combination with other planets.  As the upper panel 
of Fig.~1 shows, the planet orbiting 47UMa would almost never fall 
in the standard lensing 
zone of a $1.0M_{\odot}$ primary, but would have a probability 
$\le 50\%$ of falling within a 
slightly extended zone defined by $0.5$--$2.0 R_E$ simultaneously 
with other planets orbiting over a wide range (middle panel of Fig.~1).  
This distinction is important 
since, as GS have shown, the probability of Jovian-mass 
detection in high magnification events remains high even in this 
expanded zone.

The frequency with which multiple planets will reveal themselves 
in high amplification events depends of course on their actual 
frequency and the distribution of their orbital radius and mass.  
Consider a familiar system of a Jupiter and Saturn orbiting 
a $1.0M_{\odot}$ primary.  
Convolving the detection efficiencies of GS as a function of 
projected separation $b$ with the likelihood that Jupiter 
would have that $b$ simultaneously with Saturn falling 
in the extended $0.5$ -- $2.0 R_E$ lensing zone, we find 
that $\sim12\%$ of events with $u_{min} < 0.1$ would reveal the 
existence of the multiple planets. 
Since events with $u_{min} <0.1$
constitute $\sim10\%$ of all detected events, intense monitoring 
of 100 events per year could be expected to yield  
$\sim1$ multiple-planet lensing event per year, if all Galactic lenses
have planetary systems like our own Solar System.

\section{Single, Double and Triple Lenses}

In this section, we briefly review and apply the formalism needed for
calculating the magnification resulting from single, double and
triple lens configurations.  (For a more comprehensive review, see 
Schneider \& Weiss 1986, Paczy\'nski 1996, and references therein.)  

Consider a source with projected position $(\xi,\eta)$.  Following
Witt (1990), we write this in complex coordinates as
$\zeta=\xi+i\eta$. Lensing is the mapping from the source position
$\zeta$ to the image positions $z=x+iy$ given by 
the lens equation, which for $N$ point masses is (Witt 1990):
$$
\zeta=z + \sum_k^N {{\epsilon_k}\over{\bar{z}_k-\bar{z}}},
\eqno(3.1)
$$
where $z_k$ is the (complex) coordinate of mass $k$, $\epsilon_k$ is
the fractional mass of lens component $k$, and 
all distances are in units of the angular Einstein ring, 
$$
\theta_E\equiv\left[{  4GM \, D_{LS} \over D_{OL} \, D_{OS} \, c^2 }\right]^{1/2}={R_E\over D_{OL}}.
\eqno(3.2)
$$
The magnification $A_j$ of each image $j$ is given by the 
determinant of the Jacobian of the mapping (3.1), 
evaluated at that image position,
$$
A_j = \left.{1\over{ |{\rm{det}} J|}}\right|_{z=z_j},\, \, \,
{\rm{det}}J = 1 - {{\partial\zeta}\over{\partial{\bar z}}}
\overline{{{\partial\zeta}\over{\partial{\bar z}}}}.
\eqno(3.3)
$$
In microlensing, the images are unresolved and the total magnification
is given by the sum of the individual
magnifications, $A = \sum_j A_j$. The set of source positions for which the 
magnification is formally infinite, given by the condition 
${\rm{det}} J = 0$, defines a set of closed curves called caustics. 
For the remainder of the discussion, we will label the most massive 
(or only) component of the lens as $1$ and define the origin as $z_1=0$.  

For the single lens ($N=1$) case, the positions and
magnifications of the two resulting images can be found analytically 
and their total magnification is
$$
A_0 = { {u^2+2}\over {u (u^2+4)^{1/2}}},
\eqno(3.4)
$$
where $u \equiv |\zeta|$.  
For $u\rightarrow 0$, $A_0\rightarrow \infty$, and the point $u=0$
defines the caustic in the single lens case.
For rectilinear motion, $u=[ (t-t_0)^2/t_E^2 +u_{min}^2]^{1/2}$,
where $t_0$ is the time of maximum magnification, $u_{min}$ is the
minimum impact parameter, and $t_E$ is the time scale of the event,
$$
t_E\equiv {R_E \over v_\perp}=60\, {\rm days} \left( {M\over
    M_{\odot}}\right)^{1/2} \left( { v_\perp\over 100\, {\rm km\,
      s^{-1}}}\right)^{-1}.
\eqno(3.5),
$$
where $v_\perp$ is the transverse velocity of the lens relative to the
observer-source line-of-sight.    A single lens 
light curve is then given by $F=A_0 F_0$, where $F_0$ is the unlensed
flux of the source, and is a function of four parameters: 
$t_0$, $t_E$, $u_{min}$, and $F_0$.

For a double lens ($N=2$), Eq.~(3.1) is equivalent to a
fifth-order complex polynomial in $z$.  
The solution yields three or five images, with the number of images 
changing by two as the source crosses a caustic.  
A binary lens generates one, two, or three caustic curves, in all cases 
separate and non-intersecting.  
The light curve of a binary lens is a function of
seven parameters: the four parameters describing the single lens case,
with the additional parameters $b$, the separation of the lenses in 
units of $\theta_E$, $q$, the mass ratio of the system, and $\theta$, the
angle of the source trajectory with respect to the binary axis. 

Adding a third component to the lens system increases the complexity
enormously, in particular the caustics can exhibit self-intersection
and nesting.  Eq.~(3.1) is now equivalent to a (rather cumbersome)
10th-order polynomial in $z$.  There are thus a maximum of ten images,
and a minimum of four images, with the number of images changing by 
a multiple of two (Rhie 1997) as the source
crosses a caustic.  Although in principle 
the image positions can be found by solving the 10th-order equation, 
the process is slow and one cannot easily account for finite source sizes.  
We therefore adopt here the alternative approach of inverse ray-shooting.  
We first sample the image plane very densely and then bin the 
source positions (Eq.~3.1) in the source plane.  The ratio of the
resulting density in the source plane to the density in the image
plane gives the magnification, and repeating for all source positions 
generates a magnification map in the source plane.
We then convolve this map with different source profiles to produce  
magnifications appropriate for finite sources (Wambsganss 1996). 
Linear interpolation of the final map yields light curves for a 
particular source trajectory.  A triple lens light curve
is a function of ten parameters: the four single lens parameters, 
the separations and mass ratios $b_1$, $b_2$ and $q_1$, $q_2$, 
the angle of the source trajectory $\theta$, and $\Delta \theta$, the 
angle between the position vectors of the two companions. 

For binary and triple systems with small mass ratio(s), 
most source positions have magnifications that are nearly identical 
to that of a single lens, $A_0$.
It is thus useful to define the fractional deviation, 
 $\delta \equiv {(A -A_0)/ A_0}$, 
where $A$ is the true (binary or ternary lens) magnification.  

\section{Illustrating the Effect of Multiple Planets}

An exhaustive study of triple lenses, which
would necessitate a exploration of the $q_1$, $q_2$, $b_1$, $b_2$, and
$\Delta \theta$ parameter space, is quite beyond the scope of this
{\it Letter\/}.   
However, in order to illustrate the 
effect that a third lens would have on typical light curves we 
consider Jovian planets orbiting stars common in the Galaxy. 
Fixing $b_1=1.2$ and $q_1=0.003$, corresponding to a $M_J$ planet 
orbiting a $0.3M_{\odot}$ primary, or a $3M_J$ planet 
orbiting a $M_{\odot}$ primary, we vary the
 $b_2$ of the second planet with mass ratio $q_2=0.001$, corresponding
to a Saturn-mass planet (for the $0.3M_{\odot}$ primary) or a Jupiter-mass
planet (1.0$M_{\odot}$ primary). We concentrate on only those
source positions $|\zeta|\le 0.1$ for which the planets have a 
significant cooperative effect.   The panels of Fig.~2 show the 
magnification pattern for separations of $b_2=0.8, 1.0, 1.2$, and
 $1.4$ and relative angles 
 $\Delta \theta=0, 60^{\circ}, 120^{\circ}$, and $180^{\circ}$.  
For comparison, we also show the magnification pattern when only 
the planet of mass ratio $q_1=0.003$ is present.  For these maps,
we have adopted a uniformly-bright source with radius appropriate to a
main-sequence star, $\rho=\theta_*/\theta_E \simeq 0.003(R_*/R_\odot)$,
where $\theta_*$ and $R_*$ are the angular and physical sizes of the
source, and we have assumed $D_{OL}=6\,{\rm kpc}$, $D_{OS}=8\,{\rm
  kpc}$, and $M=0.3M_{\odot}$.

Note that the case with $b_2=1.2$ and $\Delta \theta=0$ is
completely degenerate with those from 
a single planet of mass ratio $q=q_1+q_2=0.004$.  
While for other configurations the magnification patterns with and without 
the second planet appear dramatically different, it should
be kept in mind that what one actually measures are light curves,
one-dimensional cuts through these diagrams.  
Light curves are shown in 
Fig.~3, with source radii of $\rho=0.003$ and $\rho=0.01$, for the 
sample source trajectory indicated in Fig.~2.   
Some geometries give rise to light curves that deviate dramatically 
from the case with only one planet, but those with $\Delta \theta =0$ 
have shapes that are very similar to single-planet lensing, though 
with larger amplitude and duration.  In other words, some geometries 
with two planets (i.e., those with $\Delta \theta \sim 0$ or 
$\sim 180^{\circ}$) will give rise to light curves that are degenerate 
with single planets of larger mass ratios.  Furthermore, note 
that regions of positive and negative
deviations are more closely spaced when two planets are present.
When finite source effects are considered, the overall amplitude
of the multiple planet 
anomaly will thus be suppressed.  Examples can be seen in the $b_2=1.2$ and
 $0.8$ and $\Delta\theta=180^{\circ}$ panels of Fig.~3, where 
for source radius $\rho=0.01$ the amplitude of the anomaly is smaller 
for the double planetary system than the single-planet system, while for
 $\rho=0.003$ the amplitudes are similar.  Overall detection
probabilities may thus be lower for high magnification events when
multiple planets and large sources are considered.

\section{Implications and Conclusion}

In this {\it Letter\/}, we have demonstrated that: 
(1) the probability of two
planets having projected separations that fall in the ``standard lensing zone''
($0.6 < b < 1.6$) is quite high, $\sim 1-15\%$ for planets with true separations
corresponding to Jupiter and Saturn orbiting stars of typical mass; 
(2) the influence of multiple planets in and somewhat beyond 
the standard lensing zone can be profound for high magnification 
events ($u_{min}<0.1$,) however 
(3) for some geometries, the magnification pattern and resulting 
light curves from multiple planets are qualitatively degenerate
with those from single-planet lensing, and 
(4) for high magnification events, finite source effects
are likely to suppress more substantially the amplitude of
multiple planet deviations than single planet deviations.  

Given these results, it would appear that the effects of
multiple planets on the detection and characterization of planetary
systems warrant future study.  
All previous theoretical studies have calculated microlensing planet 
detection sensitivities either by ignoring multiple planets or
by treating each planet independently.  
For high impact parameter events (low magnification), 
this is probably a fair assumption, 
but as the magnification maps in Fig.~2 illustrate 
detection probabilities will need to be revised for 
small impact parameters (large magnification).  
The sense of revision will likely depend on finite source effects.
It is also likely that for some geometries serious degeneracies exist between
light curves arising from multiple and single planet high magnification
events; these degeneracies are above and beyond those present in the
single planet case discussed by Griest \& Safizadeh (1998).  This
possible degeneracy is
especially pertinent in light of the fact that the conditional
probability of having two planets in the lensing zone is substantial.  
Thus, the interpretation of any given high magnification event may be
difficult: the degeneracies should be characterized and
their severity determined in order to have a clear understanding of
the kinds of systems whose parameters can be unambiguously determined
from the deviations.  
Finally, the calculation of planet detection efficiencies for high 
magnification events should consider multiple planets  
in order to be able to reliably convert the observed frequency of 
planetary deviations into a true frequency of planetary systems.

\acknowledgements

We would like to thank the members of the PLANET collaboration,
and especially Martin Dominik, for comments on
an earlier version of this Letter. This work was supported by grant 
AST-95-3061 from the NSF, the Kapteyn Astronomical Institute and
NWO grant 781-76-018.

\newpage

\begin{figure}
\postscript{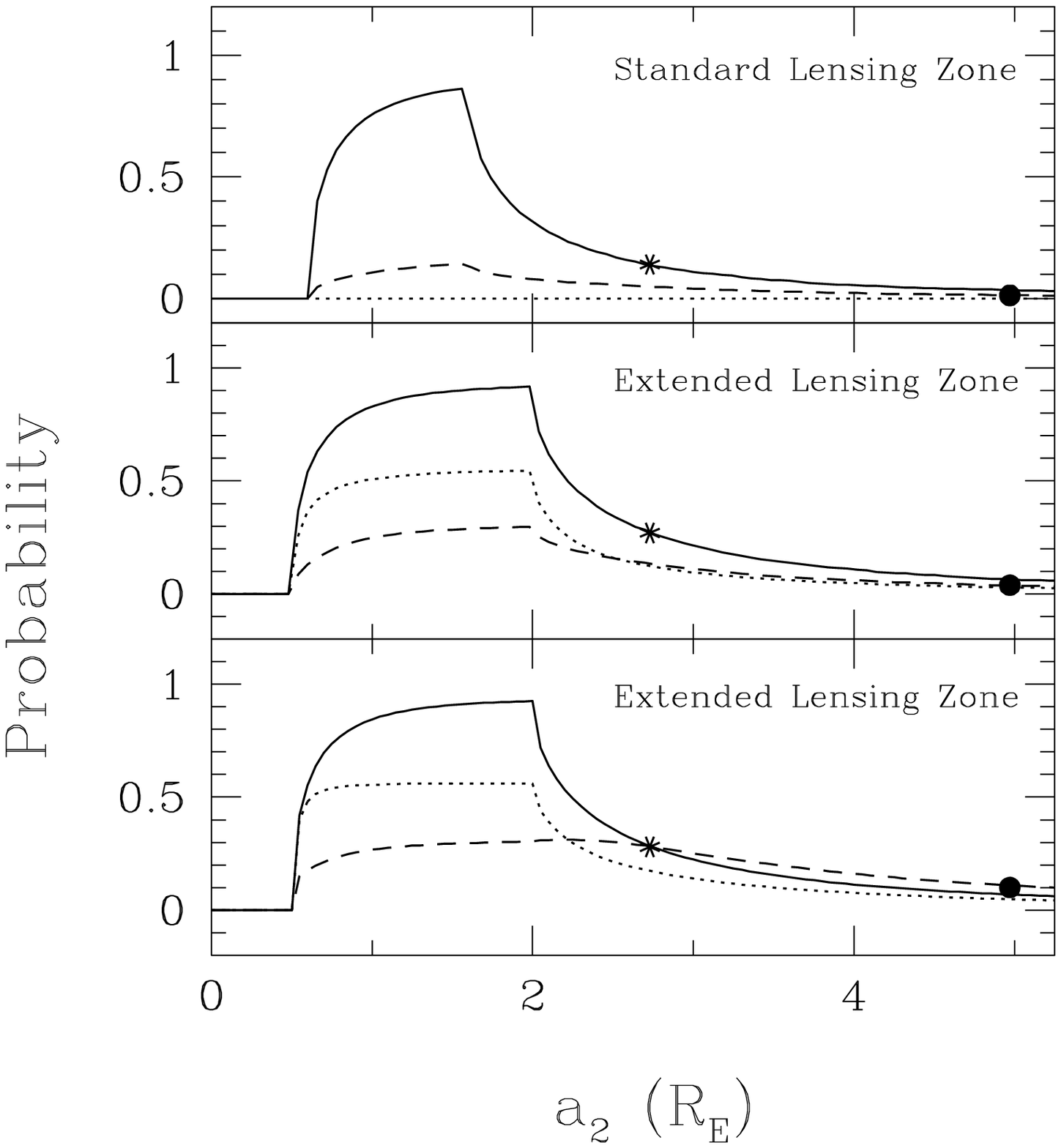}{0.9}
\caption{
Upper Panel: The probability that two planets with true separations
$a_1$ and $a_2$ (in units of the Einstein ring radius $R_E$ of the system)
will simultaneously have projected separations, 
$b_1$ and $b_2$, lying in the standard ``lensing
zone,'' defined as $0.6 < b < 1.6$.  The probability is shown
as function of $a_2$, for $a_1=1.5$ (solid, appropriate to Jupiter
orbiting a $1.0M_{\odot}$ primary), $a_1=0.6$ (dotted, appropriate
to 47UMa) and 
$a_1=2.7$ (dashed, appropriate to Jupiter orbiting a $0.3M_{\odot}$ primary).  
The probability for two planets with true
separations of Jupiter and Saturn are indicated for a solar mass
primary (star) and a $0.3M_{\odot}$ primary (dot).
Middle Panel: Same as upper panel, but for the extended ``lensing zone,'' 
$0.5 < b < 2.0$.  Lower Panel: The conditional probability that 
both $b_1$ and $b_2$ have projected separations in the extended ``lensing
zone,'' given that either $b_1$ or $b_2$ satisfies this criterion.
}
\end{figure}
\newpage
\begin{figure}
\postscript{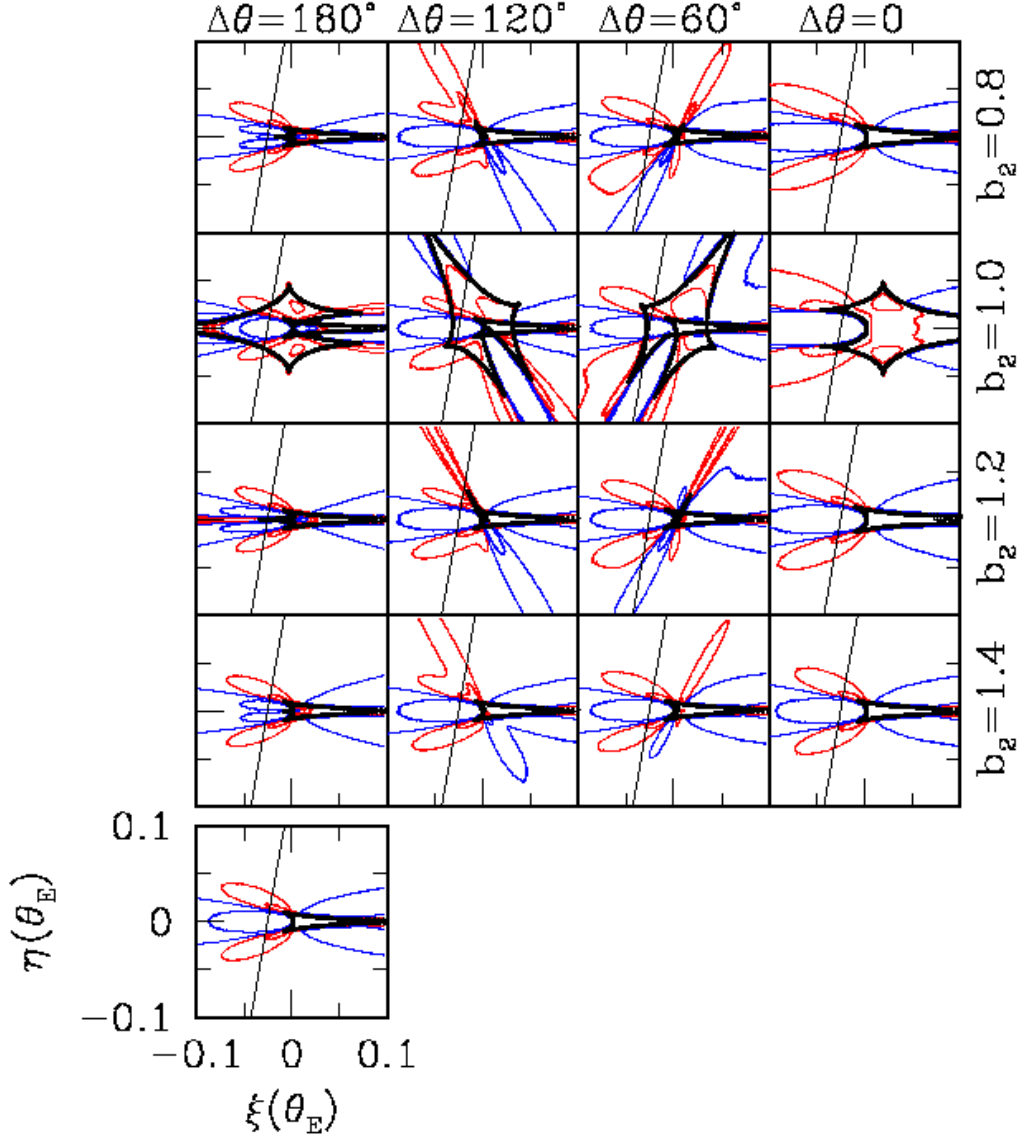}{0.9}
\caption{
Contours of constant fractional deviation $\delta$ from the
single mass lens magnification, as a function of source position
$(\xi,\eta)$ in units of the angular 
Einstein ring, $\theta_E$.  The parameters of planet 1 are held fixed
at $q_1=0.003$, $b_1=1.2$, while the projected separation 
$b_2$ and the angle between the axes, $\Delta\theta$, are varied
for a second planet with $q_2=0.001$.  The offset panel is the case
when only planet 1 is present.  Contours are $\delta=\pm 5\%$ (light lines)
and $\pm 20\%$ (bold lines). Positive contours are red, negative contours
are blue.  The caustics ($\delta=\infty$) are shown in black.
A trajectory with minimum impact parameter 
$u_{min}=0.025$ and angle $\theta=260^{\circ}$ with
respect to axis 1 is shown.}
\end{figure}
\newpage
\begin{figure}
\postscript{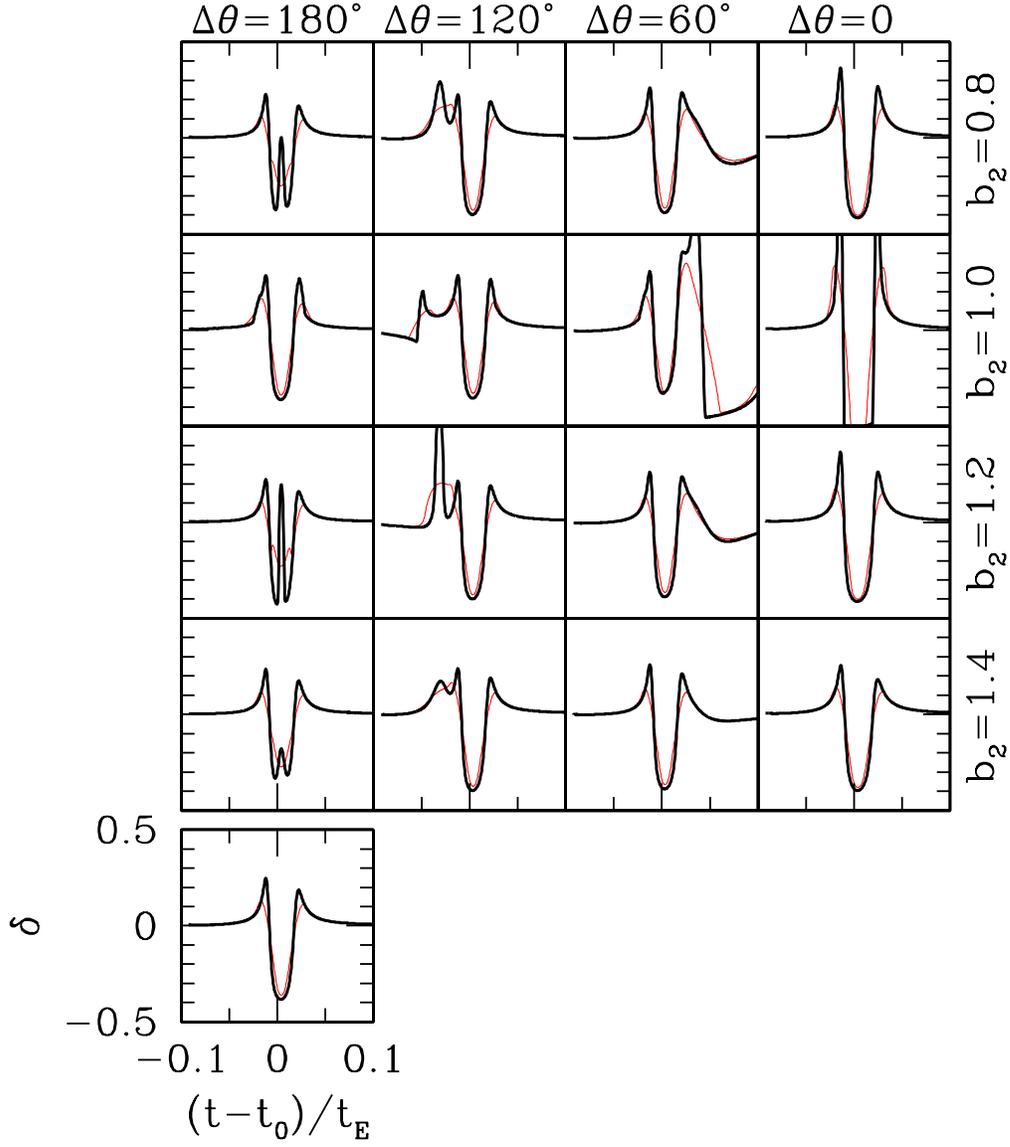}{1.1}
\caption{
The fractional deviation 
$\delta$ from a single mass lens 
as a function of time for the trajectories
shown in Fig.~2.  The black line is for a source of radius 
$\rho=0.003$ in units of $\theta_E$; the red line is for $\rho=0.01$.}
\end{figure}
\newpage
\end{document}